\documentclass[journal=ancac3,manuscript=article,layout=twocolumn]{achemso}

 \pdfoutput=1
\usepackage[version=3]{mhchem} 
\usepackage{graphicx}
\usepackage{hyperref}
\usepackage{amsmath}

\author{Hema C.P. Movva}
\email{hemacp@utexas.edu}
\author{Amritesh Rai}
\author{Sangwoo Kang}
\author{Kyounghwan Kim}
\author{Babak Fallahazad}
\affiliation{Microelectronics Research Center, The University of Texas at Austin, Austin, TX 78758, USA}
\author{Takashi Taniguchi}
\author{Kenji Watanabe}
\affiliation{National Institute of Materials Science, 1-1 Namiki, Tsukuba, 305-044, Japan}
\author{Emanuel Tutuc}
\author{Sanjay K. Banerjee}
\affiliation{Microelectronics Research Center, The University of Texas at Austin, Austin, TX 78758, USA}

\title[WSe2 FETs]{High-Mobility Holes in Dual-Gated WSe$_2$ Field-Effect Transistors}


\begin{document}

\begin{abstract}
We demonstrate dual-gated \textit{p}-type field-effect transistors (FETs) based on few-layer tungsten diselenide (WSe$_2$) using high work-function platinum source/drain contacts, and a hexagonal boron nitride top-gate dielectric. A device topology with contacts underneath the WSe$_2$ results in \textit{p}-FETs with $I_{ON}$/$I_{OFF}$ ratios exceeding 10$^7$, and contacts that remain Ohmic down to cryogenic temperatures. The output characteristics show current saturation and gate tunable negative differential resistance. The devices show intrinsic hole mobilities around 140 cm$^2$/Vs at room temperature, and approaching 4,000 cm$^2$/Vs at 2 K. Temperature-dependent transport measurements show a metal-insulator transition, with an insulating phase at low densities, and a metallic phase at high densities. The mobility shows a strong temperature dependence consistent with phonon scattering, and saturates at low temperatures, possibly limited by Coulomb scattering, or defects.
\end{abstract}


The isolation of graphene, and study of its exceptional properties has triggered an interest in several other two-dimensional (2D) layered materials\cite{vdWHet}, semiconducting transition metal dichalcogenides (TMDs)\cite{TMDChem, JariwalaReview} being one of them. In contrast to graphene's zero band-gap, semiconducting TMDs have a large (1-2 eV) band-gap, making them potentially useful for future electronic devices requiring high $I_{ON}$/$I_{OFF}$ ratios. The diverse variety of semiconducting TMDs such as molybdenum disulfide (MoS$_2$)\cite{MoS2Kis, MoS2Das}, molybdenum diselenide (MoSe$_2$)\cite{MoSe2Larentis}, tungsten disulfide (WS$_2$)\cite{WS2Jena}, tungsten diselenide (WSe$_2$)\cite{WSe2Fang, WSe2MET, WSe2Das, WSe2Podzorov}, each having its own thickness dependent electronic band-structure, provide a wide choice for specific use in optoelectronics, low-power, and/or high-performance device applications\cite{WSe2pn, WSe2MET, MoS2Das}. In addition, the coupled spin and valley degrees of freedom, and massive charge carriers in TMDs result in a wealth of novel electrical, and optoelectronic phenomena\cite{MoS2valley, MoS2spin} that can be exploited for the development of alternative device architectures\cite{BISFET, TMDBoson}. To date, MoS$_2$ has received the most attention among all TMD field-effect transistors (FETs), with the devices exhibiting \textit{n}-type conduction\cite{MoS2Kis, MoS2Das, MoS2MIT, MoS2Intrinsic, MoS2Band, MoS2WangMIT}. It is equally important to explore \textit{p}-type TMDs, in order to realize a practical TMD-based post-silicon CMOS architecture. One TMD that has attracted significant attention for \textit{p}-FETs is WSe$_2$, with early reports of bulk-WSe$_2$ FETs showing hole mobilities approaching 500 cm$^2$/Vs\cite{WSe2Podzorov}. Subsequently, few-layered WSe$_2$ FETs have also been demonstrated with high $I_{ON}$/$I_{OFF}$ ratios, and hole mobilities\cite{WSe2Fang, WSe2Das, WSe2hBN, WSe2Pradhan, WSe2mono}. While MoTe$_2$\cite{MoTe2FET}, and 2D black phosphorus\cite{BlackPFET} have also been reported to show \textit{p}-type conduction, these materials are less stable in ambient conditions. The high thermal, and environmental stability, and well-developed materials science of WSe$_2$ make it very attractive as a channel material for 2D \textit{p}-FETs\cite{WSe2Podzorov}.

Creating low resistance, Ohmic contacts has been a major challenge limiting study of the intrinsic properties of TMDs. Most metal contacts to TMDs form Schottky barriers, resulting in large series resistances which degrade even further at low temperatures\cite{MoS2Das, WSe2MET, WSe2Das}. Considerable research effort has been put into addressing this problem, using techniques such as metal work-function tuning\cite{MoS2Das, WSe2MET, MoOxContact}, contact annealing\cite{MoS2Intrinsic, MoS2MIT}, graphene contacts\cite{MoS2Graph, WSe2EDLT, WSe2hBN}, electrical double layer (EDL) structures\cite{WSe2mono, WSe2EDLT, WSe2hBN}, and doped source/drain contacts\cite{WSe2Fang}. While improving the contacts, these techniques however have several limitations like (i) processing constraints, and instability of low work-function metals\cite{MoS2Das, WSe2MET}, (ii) unintentional doping during contact annealing\cite{MoS2Intrinsic}, (iii) slow response speed of EDL structures, preventing their use in FETs\cite{WSe2mono}, and (iv) instability of surface charge transfer dopants in air\cite{WSe2Fang}. Whereas graphene contacts result in efficient electron injection in MoS$_2$\cite{MoS2Graph}, the large band offsets between the Dirac point of graphene, and the conduction, and valence bands of WSe$_2$\cite{WSe2Kyoung} necessitate additional doping of the graphene for efficient carrier injection\cite{WSe2hBN, WSe2EDLT}. Moreover, these approaches are primarily directed towards back-gated FET geometries, which are of limited use in practical circuits. There is a need to develop top-gated FET structures in order to enable independent control of multiple FETs on the same substrate towards large scale device integration. Furthermore, an air-stable, low temperature compatible contact scheme is imperative for a systematic investigation of the nature of charge transport in WSe$_2$.

In this work, we use high work-function platinum (Pt) as the contact metal for efficient hole injection into the valence band of WSe$_2$. By using a device topology with the Pt contacts underneath the WSe$_2$, and a pristine hexagonal boron nitride (hBN) top-gate dielectric, we realize dual-gated FETs  with contacts that are Ohmic down to cryogenic temperatures. We demonstrate that this contact scheme is optimized for top-gated FET operation, with the back-gate serving as an additional knob to fine-tune the FET characteristics. Top-gated transfer characteristics show $I_{ON}$/$I_{OFF}$ ratios exceeding 10$^7$, and hole mobilities around 140 cm$^2$/Vs at room temperature in three/four-layer WSe$_2$. The output characteristics exhibit current saturation, and a negative differential resistance. Temperature-dependent transport measurements reveal a metal-insulator transition, indicating high device quality. The mobility shows a strong temperature dependence at high temperatures, indicative of phonon dominated transport in this regime. At low temperatures, the mobility saturates, approaching up to 4,000 cm$^2$/Vs, possibly limited by Coulomb scattering, or defects\cite{WSe2hBN, MoS2Intrinsic}.

\section{Results and discussion}

\begin{figure*}
  \includegraphics{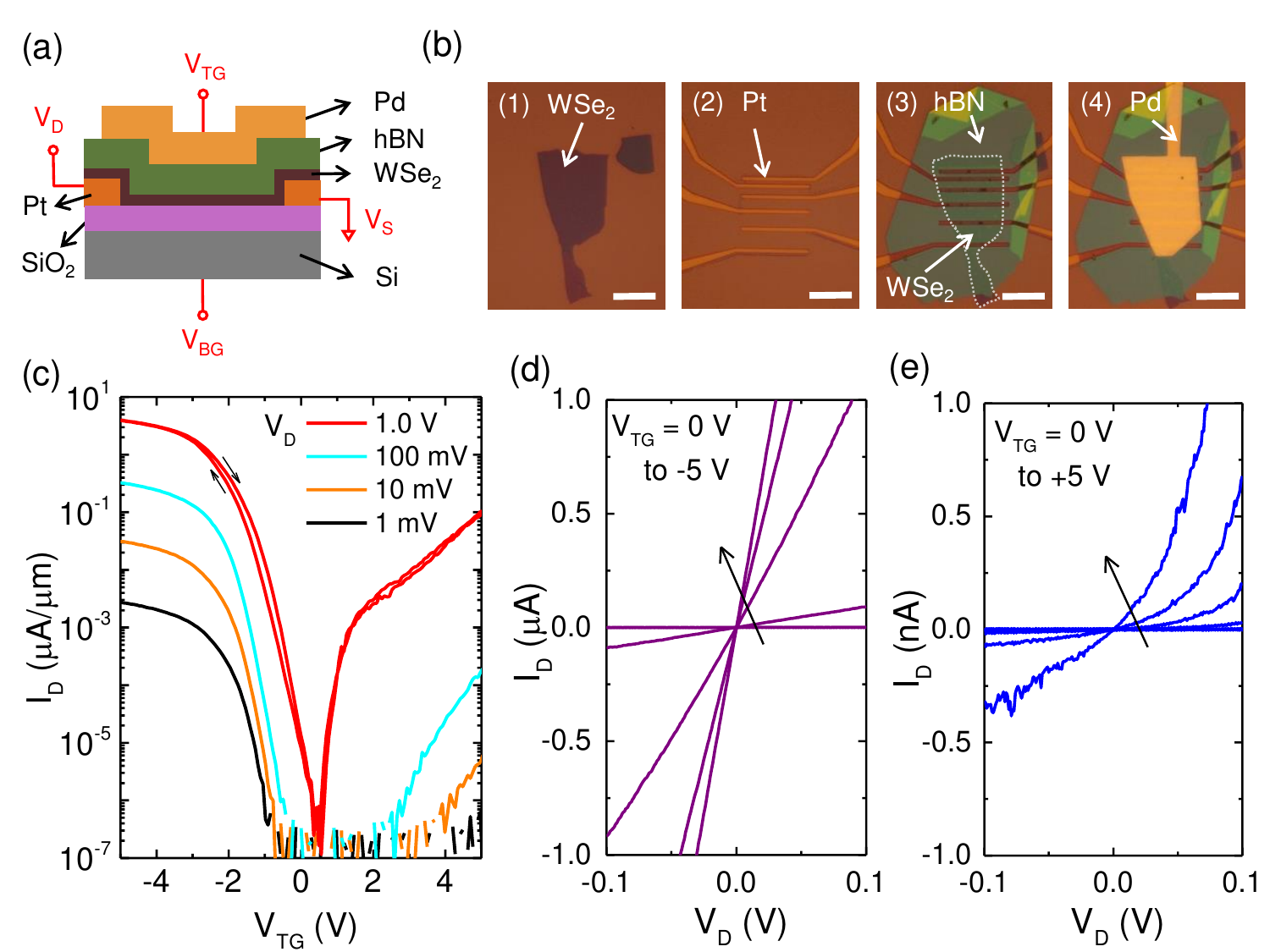}
  \caption{(a) Schematic of a dual-gated WSe$_2$ FET with Pt contacts underneath the WSe$_2$, an hBN top-gate dielectric, and a Pd top-gate. The biasing scheme is shown in red. (b) Optical micrographs during the fabrication process show (1) a three/four-layer exfoliated WSe$_2$ flake on SiO$_2$/Si, and (2) pre-patterned Pt contacts on a separate SiO$_2$/Si substrate. (3) An hBN flake is used to pick-up the WSe$_2$, and is transferred onto the Pt contacts, followed by (4) patterning of a local Pd top-gate. The scale bars are 10 $\mu$m (c) Transfer characteristics of the FET ($L$ = 6 $\mu$m) at different $V_{D}$, showing \textit{p}-type conduction at $V_{D}$ as low as 1 mV, and an $I_{ON}$/$I_{OFF}$ ratio $>$ $10^7$ at $V_{D}$ = 1 V. Low-bias output characteristics of the FET at different $V_{TG}$ for the (d) \textit{p-branch}, and (e) \textit{n-branch} show Ohmic, and Schottky contacts, respectively. The back-gate is grounded for all measurements.}
  \label{fig1}
\end{figure*}

We use exfoliated WSe$_2$ flakes derived from commercially available crystals as the source material for fabricating the FETs in this work. Three/four-layer WSe$_2$, and 15-20 nm hBN flakes are identified using optical contrast, Raman spectroscopy, and photoluminescence measurements (S1 in Supporting Information). A polymer-coated silicone stamp\cite{OneDcontact} is used to assemble, and transfer a stack of hBN/WSe$_2$ on to pre-patterned Cr/Pt electrodes on a SiO$_2$/Si substrate. Subsequently, a local palladium (Pd) top-gate is patterned, resulting in a device structure as shown in Figure~\ref{fig1}(a). Optical micrographs during the fabrication process are shown in Figure~\ref{fig1}(b), and described in detail in the Materials and Methods section.

We chose Pt due to its high work-function ($\Phi_{M}$ $\sim$ 6.0 eV)\cite{PtWorkF}, which places its Fermi level below the valence band edge of WSe$_2$ ($\chi_{WSe_{2}}$ + $E_{g}$ $\sim$ 5.5 eV)\cite{WSe2Kyoung}.This band alignment is intuitively expected to result in Ohmic \textit{p}-type contacts. Choice of appropriate work-function metals has been successfully employed in the past to optimize carrier injection in TMD FETs. Low work-function scandium contacts were found to result in efficient electron injection in MoS$_2$\cite{MoS2Das}, and indium was used for low-resistance \textit{n}-type contacts to WSe$_2$\cite{WSe2MET}. However, Fermi level pinning at the metal-TMD interface has also been found to strongly impact the contacts, resulting in non-trivial behavior for some metal-TMD combinations.\cite{MoS2Das, WSe2MET} Furthermore, the metal-TMD interface is highly sensitive to the processing environment such as, vacuum conditions in the deposition chamber, deposition rate, metal topography, \textit{etc}. These variations can potentially affect the TMD electronic structure, metal crystallinity, and in turn, the metal work-function\cite{MoS2MetCont}, resulting in wide variations in FET characteristics among reports in literature\cite{MoS2Das, WSe2MET, WSe2Pradhan}. By choosing an inert contact metal like Pt, and by decoupling the metal deposition step from creating the actual metal-TMD contact, we can potentially eliminate some of these uncertainties. Direct deposition of Pt on WSe$_2$ as a top-contact is impractical due to its poor adhesion, whereas, using an adhesion layer such as chromium reduces the effective metal work-function at the contact interface. Our strategy of back-contacts circumvents this problem, since the Pt electrodes can now be deposited with an appropriate adhesion layer at the bottom without affecting the top surface work-function. The transferred WSe$_2$ would contact the Pt top-layer, whose high work-function would still be preserved. Finally, the choice of hBN as the top-gate dielectric is motivated by its ultra-flat surface, which has been shown to reduce extrinsic impurity scattering in TMDs\cite{WSe2hBN, MoS2Graph, WSe2EDLT}.

We evaluate the effectiveness of our back-contact scheme by performing gate-dependent transport measurements. Figure~\ref{fig1}(c) shows the top-gated transfer characteristics of a WSe$_2$ FET with an 18 nm hBN top-gate dielectric at different values of drain bias ($V_{{D}}$), with the back-gate grounded. The biasing scheme is shown in Figure~\ref{fig1}(a). The top-gate bias ($V_{TG}$) is applied to the Pd top-gate, the back-gate bias ($V_{BG}$) to the highly-doped Si substrate, and the source terminal ($V_{S}$) is grounded. Even at a low $V_{D}$ = 1 mV, the drain current ($I_{D}$) is found to increase with increasing negative $V_{TG}$, and an insulating behavior is observed for positive $V_{TG}$. This indicates predominant hole transport in the WSe$_2$ for negative $V_{TG}$, hereafter referred to as the ``\textit{p-branch}". For larger $V_\text{D}$, however, we observe an increase of $I_{D}$ even with increasing positive $V_{TG}$ (hereafter called the ``\textit{n-branch}"), symptomatic of emergent electron conduction in this regime. However, while $I_{D}$ increases proportionally with $V_{D}$ for the \textit{p-branch}, the behavior is highly non-linear for the \textit{n-branch}. The overall behavior hints at Ohmic, and Schottky nature of the Pt back-contacts to the valence, and conduction bands of WSe$_2$, respectively. Similar ambipolar characteristics reported for WSe$_2$ FETs with conventional top-contacts\cite{WSe2hBN, WSe2Das, WSe2EDLT, WSe2mono} confirm that the transferred WSe$_2$ does indeed form a good electrical contact with the Pt electrodes. Evidence of clear subthreshold, and insulating regimes, along with a high $I_{ON}$/$I_{OFF}$ ratio over 10$^7$ (at $V_{D}$ = 1 V) further demonstrate that the integrity of the WSe$_2$ is maintained during, and after transfer. The intrinsic nature of WSe$_2$ is evident from its insulating state around $V_{TG}$ = 0 V, indicating no unintentional doping from the fabrication process. Negligible hysteresis in the transfer characteristics signifies minimal charge trapping, and therefore, clean interfaces in the device. A threshold voltage ($V_{T}$) of -2 V can be extracted for the \textit{p-branch} from the linear region of the transfer characteristics. The quality of the hBN dielectric is manifested in the top-gate leakage current, which remains close to the noise floor throughout the measured $V_{TG}$ range (S2 in Supporting Information).

The nature of contacts can be further verified from the low-bias output characteristics that are shown in Figures~\ref{fig1}(d), and (e). The characteristics for the \textit{p-branch} (Figure~\ref{fig1}(d)) show a symmetric, linear dependence of $I_{D}$ on $V_{D}$ for all negative values of $V_{TG}$, denoting Ohmic contacts. On the other hand, the trend for positive $V_{TG}$ (Figure~\ref{fig1}(e)) is highly non-linear, and asymmetric, indicative of Schottky contacts to the  \textit{n-branch}. It should be noted that the currents for the \textit{p-branch} are more than three orders of magnitude larger than the \textit{n-branch}, confirming that Pt is better suited to contact the valence band of WSe$_2$. The high work-function of Pt results in a large Schottky barrier to the conduction band which explains the Schottky nature of contacts to the \textit{n-branch}. Lower work-function metals such as indium, silver,\cite{WSe2MET} nickel\cite{WSe2Das}, \textit{etc}. would be preferable for contacting the \textit{n-branch} of WSe$_2$. The integration of back-side source/drain contacts with a top-gated geometry ensures unimpeded electrostatic modulation of the contact, and channel access regions by the top-gate. By contrast, in top-gated TMD FETs with top contacts, screening by the source/drain electrodes obstructs modulation of these access regions by the top-gate, and the ensuing large series resistances severely limit current injection into the channel, and degrade the FET performance\cite{MoS2Kis,WSe2Fang}. Additionally, our device structure with inert metal electrodes, and dielectrics is robust, and can be extended to other TMDs, and TMD heterostructures, where large series resistances are particularly problematic\cite{TMDHetJavey}.

\begin{figure*}
  \includegraphics{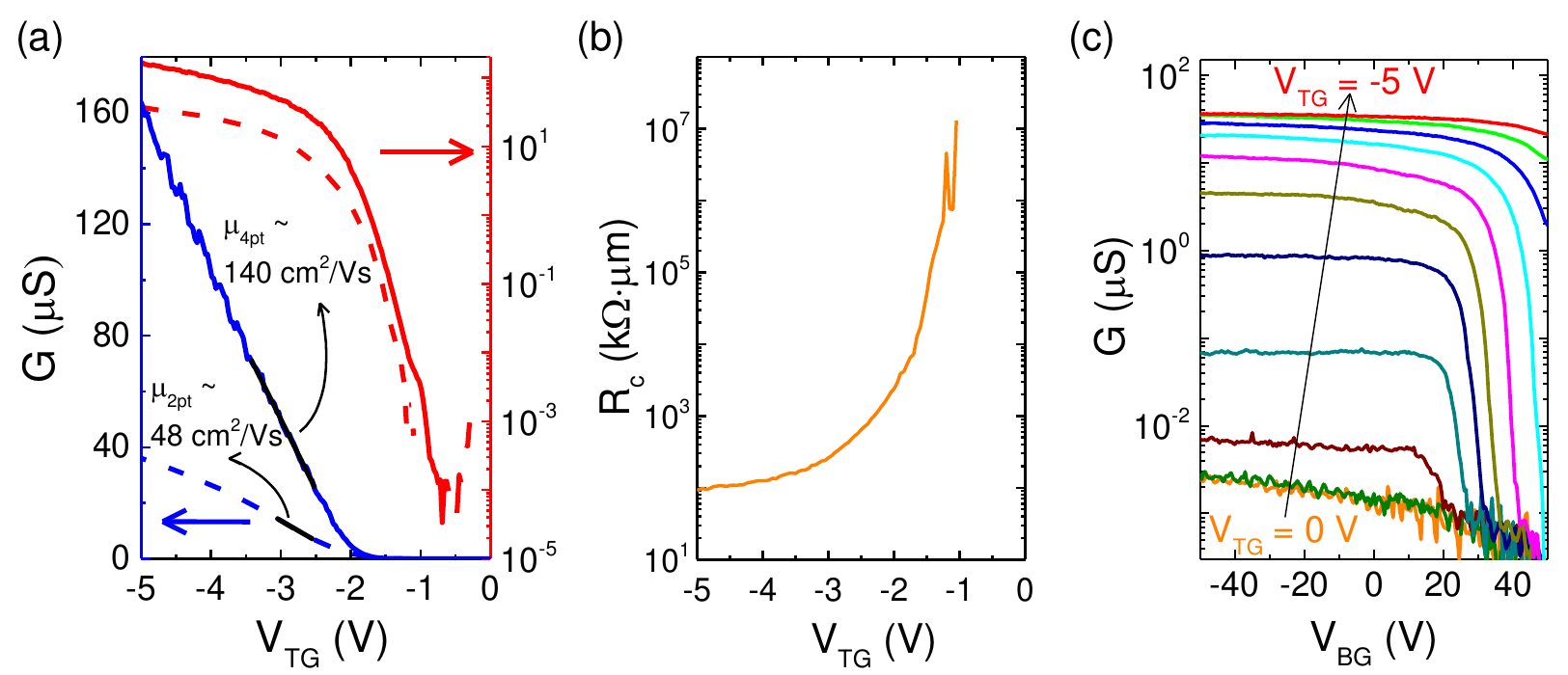}
  \caption{(a) Transfer characteristics of a WSe$_2$ FET ($L$ = 6 $\mu$m, $W$ = 12 $\mu$m) showing variation of $G_{2pt}$ (dashed lines), and $G_{4pt}$ (solid lines) as a function of $V_{TG}$, in linear (left, blue) and log (right, red) scales. The black lines show linear fits to Equation 1, resulting in $\mu_{2pt}$ = 48 cm$^2$/Vs, and an intrinsic, $\mu_{4pt}$ = 140 cm$^2$/Vs. (b) A plot of the variation of $R_{c}$ \textit{vs.} $V_{TG}$ shows $R_{c}$ reducing with $V_{TG}$ becoming more negative due to modulation of the channel access, and contact regions by the top-gate. The back-gate is grounded in (a) and (b). (c) Back-gated transfer characteristics show saturation of $G_{2pt}$ in the ON state, which increases as $V_{TG}$ becomes more negative.}
  \label{fig2}
\end{figure*}

In the following, we focus the discussion on hole transport in our devices. We use multi-terminal four-point measurements to extract the intrinsic hole mobilities, and contact resistances. Figure~\ref{fig2}(a) shows the 2-point conductance ($G_{2pt}$), and 4-point, intrinsic conductance ($G_{4pt}$) as a function of $V_{TG}$. While $G_{2pt}$ is measured as the conductance between an adjacent pair of contacts, $G_{4pt}$ is measured using the voltage drop between the same two contacts when biasing an outer set of contacts (S3 in Supporting Information). The field-effect mobility ($\mu$) is then calculated using

\begin{equation}
\mu = \frac{1}{C_{TG}} \frac{L}{W} \frac{dG}{dV_{TG}}
\label{eq1}
\end{equation}

\noindent where, $G$ is the channel conductance, $C_{TG}$ is the top-gate capacitance, and $W$, and $L$, are the width, and length of the channel, respectively. For extracting the 2-point field-effect mobility ($\mu_{2pt}$), we use $G$ = $G_{2pt}$, and for the 4-point, intrinsic field-effect mobility ($\mu_{4pt}$), we use $G$ = $G_{4pt}$. While $G_{2pt}$ has contributions from both the intrinsic channel conductance, and contact resistance, $G_{4pt}$ is a measure of only the intrinsic channel conductance. The specific contact resistance ($R_{c}$) can therefore be determined using

\begin{equation}
R_{c} = \frac{L}{2}\Big(\frac{1}{G_{2pt}} - \frac{1}{G_{4pt}}\Big)
\label{eq2}
\end{equation}

\noindent where, $R_{c}$ is normalized to the contact width, and has units of k$\mathrm{\Omega\cdot\mu}$m. For an hBN thickness of 18 nm (dielectric constant of 3.0), which corresponds to a geometric top-gate capacitance ($C_{TG}$)\cite{WSe2hBN, MoS2Intrinsic} of 150 nF/cm$^2$, $L$ = 6 $\mu$m, and $W$ = 12 $\mu$m, we extract $\mu_{2pt}$ = 48 cm$^2$/Vs, and $\mu_{4pt}$ = 140 cm$^2$/Vs. The considerably lower value of $\mu_{2pt}$ compared to $\mu_{4pt}$ is due to the detrimental effect of $R_{c}$ on $G_{2pt}$. Acting as a parasitic series resistance, $R_{c}$ reduces the effective drive voltage on the channel, thereby reducing $G_{2pt}$, and in turn, $\mu_{2pt}$. The extracted $\mu_{4pt}$ of 140 cm$^2$/Vs compares well with prior reports of intrinsic mobilities in few-layer MoS$_2$\cite{MoS2Intrinsic, MoS2Graph, MoS2MIT}, and WSe$_2$\cite{WSe2hBN, WSe2EDLT, WSe2mono}. The hole-density at $V_{TG}$ = -5 V can be extracted using

\begin{equation}
p = \frac{C_{{TG}}|(V_{{TG}}-V_{{T}})|}{e}
\label{eq3}
\end{equation}

\noindent where, $e$ is the charge of an electron, to be $p$ = 2.8$\times$10$^{12}$ /cm$^{2}$. While this value is lower than the carrier densities attainable by EDL structures\cite{WSe2EDLT, WSe2mono}, it is comparable to densities in conventional dielectric based FETs\cite{WSe2Pradhan, WSe2hBN}.

The variation of $R_{c}$ \textit{vs.} $V_{TG}$ is plotted in Figure~\ref{fig2}(b). It can be seen that $R_{c}$ reduces as $V_{TG}$ becomes more negative, asymptotically approaching $\sim$ 100 k${\Omega \cdot \mu m}$ at $V_{TG}$ = -5 V. While exhibiting a larger $R_{c}$ than values for EDL gated WSe$_2$ ($\sim$ 10 k${\Omega \cdot \mu m}$)\cite{WSe2mono, WSe2hBN}, our Pt contacts display superior low-temperature Ohmic behavior. The strong dependence of $R_{c}$ on $V_{TG}$ is consistent with previous reports of gate-tunable contact barriers at the metal-TMD interface\cite{MoS2Intrinsic, WSe2hBN, MoS2Graph}. Variation of $R_{c}$ with gate-bias is not observed in traditional MOSFETs due to their highly-doped source/drain regions\cite{Banerjee}. However, if they are undoped, as in typical TMD FETs, the gate can electrostatically modulate the contact regions, and in turn, $R_{c}$. Whereas the top-gate can efficiently modulate the contact regions in our structure, screening by the source/drain electrodes prevents modulation by the back-gate. Back-gated FET transfer characteristics are therefore severely series resistance limited, as illustrated in Figure~\ref{fig2}(c). While the overall variation of $G_{2pt}$ \textit{vs.} $V_{BG}$ is consistent with hole conduction, a pronounced saturation is observed for negative $V_{BG}$. Similar saturation of the transfer characteristics at large gate-biases has been reported in conventional top-gated FETs with top-contacts\cite{MoS2Kis,WSe2Fang,MoS2Atresh}. Due to our device geometry, the role played by the back-gate, which modulates the channel but not the contact regions, is analogous to the role played by the top-gate in conventional top-gated TMD FETs (S4 in Supporting Information). Consequently, when $V_{TG}$ = 0 V, the contact regions are highly resistive, and inhibit current flow through the channel, resulting in a low $G_{2pt}$ for all $V_{BG}$. As $V_{TG}$ is progressively made more negative, the contact regions accumulate holes, leading to a decrease of $R_{c}$, and in turn, an increase of $G_{2pt}$ in the ON state. The $G_{2pt}$ saturation is due to $R_{c}$ dominating the total channel resistance. Since $R_{c}$ can be modulated much more effectively by $V_{TG}$, $G_{2pt}$ in the ON state is highly sensitive to $V_{TG}$ as compared to $V_{BG}$. Further, the shift in $V_{BG}$ at the onset of saturation with varying $V_{TG}$ is due to the effect of dual-gating of the channel. A more negative $V_{TG}$ accumulates additional holes in the channel, thereby requiring a more positive $V_{BG}$ to deplete them. The effect is an increase of $V_{BG}$ at the onset of saturation as $V_{TG}$ is made more negative. Finally at $V_{TG}$  = -5 V, the FET remains ON throughout the measured $V_{BG}$ range. The intrinsic nature of WSe$_{2}$ is therefore the primary reason why top-gated FETs in the conventional top-contact geometry show poor characteristics\cite{WSe2Fang, WSe2hBN}. On the other hand, our structure with back-contacts is better suited for top-gated operation since it allows for efficient top-gate modulation of the contact regions.

\begin{figure}
  \includegraphics{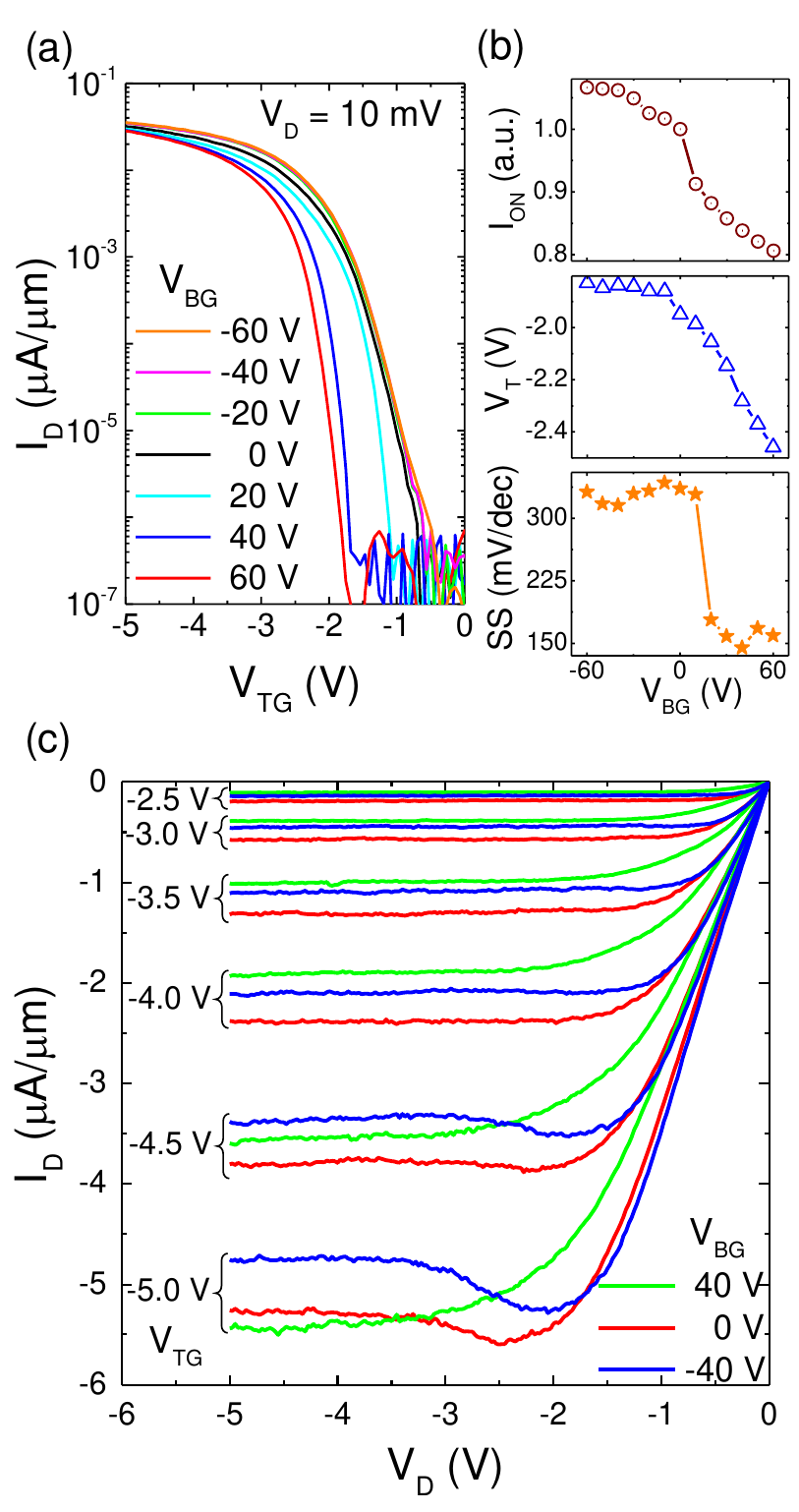}
  \caption{(a) Top-gated transfer characteristics of the FET at different values of $V_{BG}$. (b) The variation of $I_{ON}$ (top-panel), $V_{T}$ (middle-panel), and $SS$ (bottom-panel) can be understood qualitatively by considering the effect of $V_{BG}$ on the channel. (c) FET output characteristics show current saturation, and an NDR prior to the onset of saturation. The NDR magnitude changes with $V_{BG}$, increasing for $V_{BG}$ = -40 V, and gets quenched when $V_{BG}$ = 40 V.}
  \label{fig3}
\end{figure}

The back-gate can further also be used to tune the FET characteristics. Figure~\ref{fig3}(a) shows the top-gated transfer characteristics at different $V_{BG}$ values. Variation in FET parameters like ON current ($I_{ON}$), $V_{T}$, and subthreshold swing ($SS$) are apparent, and are shown in Figure~\ref{fig3}(b). A negative (positive) $V_{BG}$ increases (reduces) the hole density in the channel, leading to a $V_{T}$ shift. As a consequence, negative (positive) values of $V_{BG}$ reduce (increase) the channel resistance, and increase (reduce) $I_{ON}$. At negative values of $V_{BG}$, which increase the channel hole density, the FET turn ON is limited by the contact regions' turn ON. The relative insensitivity of the contacts to $V_{BG}$ makes $V_{T}$ insensitive to $V_{BG}$ in this regime. Two predominant regimes are evident in the $SS$, at $\sim$ 330 mV/dec for $V_{BG}$ $\leq$ 10 V, and $\sim$ 150 mV/dec for $V_{BG}$ $>$ 10 V. For $V_{BG}$ $\leq$ 10 V, the channel is accumulated with excess holes, and the $SS$ is determined by turn ON of the contact regions. In contrast, for $V_{BG}$ $>$ 10 V, the channel is populated with electrons, resulting in a steeper $SS$ of 150 mV/ dec, dictated by diffusion current from the source to drain\cite{Banerjee}. It is to be noted that the $SS$ is relatively insensitive to the carrier concentration in the channel, but depends only on the polarity of excess carriers induced by the back-gate.

Figure~\ref{fig3}(c) shows the top-gated FET output characteristics at different values of $V_{BG}$. First, there is a clear evidence of current saturation at large negative $V_{D}$ for all values of $V_{TG}$ $<$ $V_{T}$, and $V_{BG}$. Current saturation is due to channel pinch-off at the drain, similar to a conventional MOSFET. A maximum drive current of $\sim$ 5 $\mu$A/$\mu$m is obtained at $V_{TG}$ = -5 V for a long-channel device with $L$ = 6 $\mu$m, which is comparable to values reported for WSe$_2$ \textit{p}-FETs with chemically doped source/drain contacts\cite{WSe2Fang}. Higher drive currents are possible by using shorter channel lengths, and thinner top-gate dielectrics. A second feature is the negative differential resistance (NDR) behavior prior to the onset of current saturation. The NDR behavior that is commonly observed in bulk III-V FETs is due to a transferred electron mechanism, often referred to as the Gunn effect\cite{Banerjee}. Recent reports of NDR in MoS$_2$ have also been attributed to a transferred electron mechanism between satellite valleys, and/or a self-heating effect\cite{MoS2Heat, MoS2ValleyHeat}. Our devices, however, also show considerable hysteresis between the forward and reverse $I_{D}$-$V_{D}$ sweeps (S5 in Supporting Information). Both the NDR amplitude, and hysteresis are correlated, increasing for $V_{BG}$ = -40 V, and almost vanishing for $V_{BG}$ = 40 V. The NDR dependence on the $V_{BG}$ value, and polarity suggests that the vertical carrier distribution in the WSe$_2$ layer plays a key role. Application of $V_{BG}$ changes the position of the charge centroid in the WSe$_2$, with negative (positive) $V_{BG}$ shifting the holes closer to (further away from) the SiO$_2$ substrate. A real space transfer between the high mobility top layer closer to the hBN, and the low mobility bottom layer closer to the SiO$_2$ substrate could lead to an NDR behavior. The hysteresis dependence on $V_{BG}$ is further suggestive of hot carrier trapping at the WSe$_2$-SiO$_2$ interface\cite{WSe2Kyoung}, which increases (decreases) when the carriers are closer to (further away from) the SiO$_2$ substrate. It is also possible that a transferred electron mechanism could be at play, as evinced by the persistent NDR in both the forward and reverse sweeps, but the hysteresis makes it difficult to unambiguously draw this conclusion.

\begin{figure*}
  \includegraphics{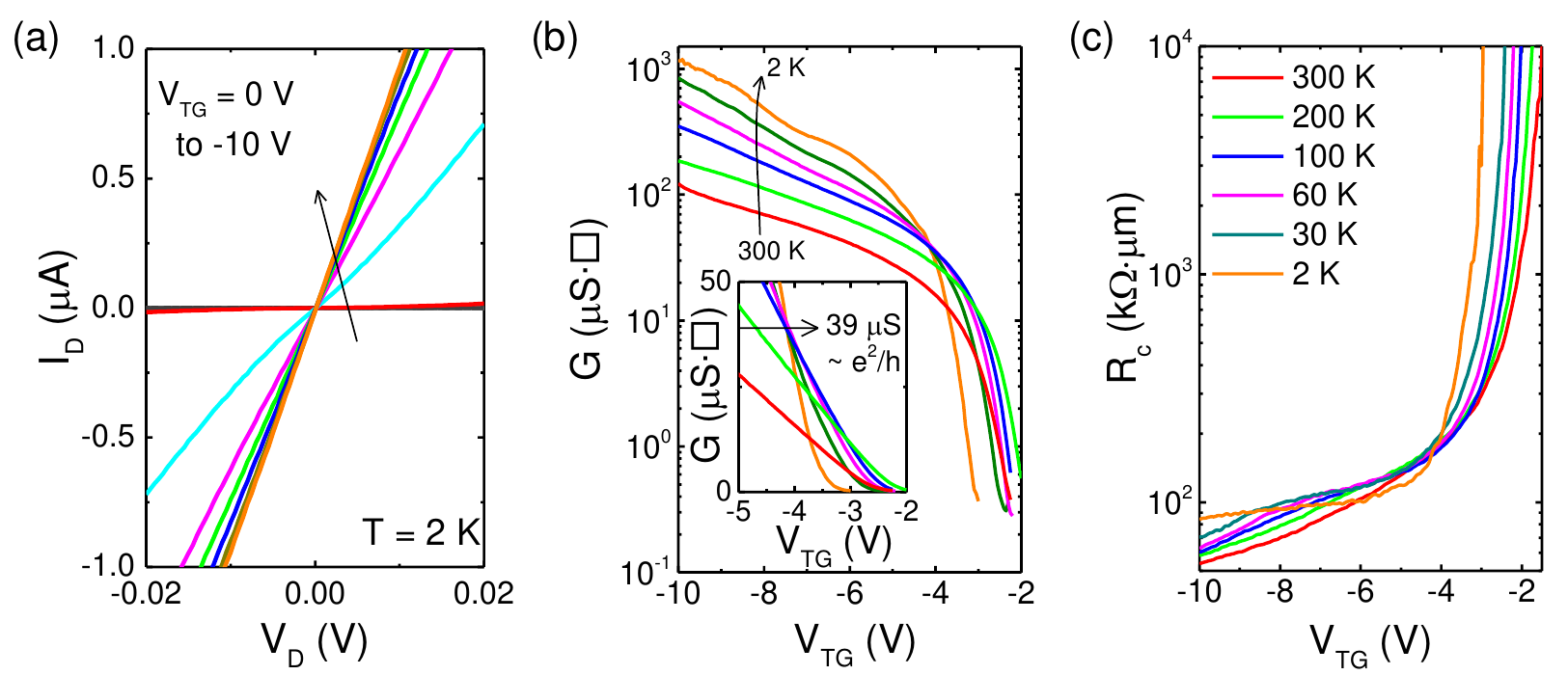}
  \caption{(a) Low-bias output characteristics showing a mostly linear $I_{D}$-$V_{D}$ behavior, indicating Ohmic nature of the Pt back-contacts even at 2 K. (b) Variation of $G$ \textit{vs.} $V_{TG}$ with temperature shows $G$ increasing with reducing temperature for $V_{TG}$ $<$ -4 V, characteristic of a metallic phase. The inset shows a close-up of the crossover, at $G_{c}$ = 39 $\mu$S ($\sim$ $e^2/h$). (c) Variation of $R_{c}$ roughly mirrors $G$, but with a weaker temperature dependence at large negative $V_{TG}$. The back-gate is grounded for all measurements.}
  \label{fig4}
\end{figure*}

We now proceed to discuss the temperature dependence of transport in our devices. The Ohmic nature of the Pt back-contacts is retained down to 2 K, as shown in Figure~\ref{fig4}(a). While the $I_{D}$-$V_{D}$ for small $V_{TG}$ shows a slight non-linearity, the behavior is more linear for large negative $V_{TG}$, where the channel has a large concentration of holes. This enables use of a standard low-frequency lock-in technique (10 nA excitation at 11.27 Hz) to measure the channel conductivity, $G$ = $G_{4pt}{\times}(L/W)$, as a function of temperature (Figure~\ref{fig4}(b)). Two distinct regimes are apparent in the temperature variation of $G$; for $V_{TG}$ $<$ -4 V, $G$ increases monotonically with decreasing temperature, and for $V_{TG}$ $>$ -4 V, $G$ does not follow a monotonic trend. The crossover between these two regimes, apparent from Figure~\ref{fig4}(b), and Supporting Information S6 suggests a metal-insulator transition (MIT), consistent with previous observations for a variety of 2D electron, and hole systems, including TMDs\cite{MoS2Intrinsic,MoS2MIT,WSe2mono,WSe2EDLT}. A close-up of the MIT point in the inset of Figure~\ref{fig4}(b) shows the crossover at a conductivity $G_{c}$ = 39 $\mu$S $\sim$ $e^2/h$. Other samples show $G_{c}$ in the same range of conductivity, albeit with slight variations (S6 in Supporting Information). To better understand the nature of MIT observed in our samples, we discuss the results using the theoretical framework developed to explain the phenomenon in a large set of 2D electron, and hole systems\cite{SarmaMIT,SarmaPerc}.

According to the scaling theory of localization, all non-interacting 2D systems exhibit an insulating ground state in the limit of zero temperature\cite{ScalingTheory}. At high carrier densities, and in samples with reduced disorder, the localization length can exceed the sample size. In this weakly localized state, the 2D system can exhibit an apparent metallic behavior, explained in terms of the temperature dependent screening of fixed charged impurities. For high sample disorder, or at low carrier densities, the system becomes strongly localized, and the temperature dependence of conductivity displays the expected insulating behavior. This crossover from a metallic weakly localized regime at high carrier densities to an insulating strongly localized regime at low carrier densities has been used to explain the MIT in 2D semiconductors\cite{SarmaMIT, MoS2MIT}. To ascertain the nature of MIT in our devices, we examine the following temperature scales: the Fermi temperature ($T_{F}$), the Bloch-Gr{\"u}neisen temperature ($T_{BG}$), and the Dingle temperature ($T_{D}$) (S7 in Supporting Information). The temperatures $T_{BG}$, and $T_{D}$ define scales associated with phonon scattering, and disorder, respectively. An unambiguous manifestation of a weak localization mediated metallic phase requires $T_{D}$ $<$ $T_{F}$ $<$ $T_{BG}$ at the crossover point, a condition which rules out phonon scattering in the metallic phase, and ensures that the disorder is sufficiently weak\cite{SarmaMIT}. For the sample of Figure~\ref{fig4}, $\mu_{4pt}$ = 2,600 cm$^2$/Vs, and the crossover carrier density ($p_{c}$) of 5.3${\times}10^{11}$/cm$^2$ at 2 K, we obtain $T_{F}$ = 29 K, $T_{BG}$ = 16 K, and $T_{D}$ = 5.0 K. Since $T_{F}$ $>$ $T_{BG}$, the temperature dependence of $G$ in the metallic phase can potentially be affected by phonon scattering in our devices. We note that in the absence of a more reliable carrier density measurement \textit{e.g.} through Hall effect, and the uncertainty in $V_{T}$, the value of $p_{c}$ could be underestimated. However, a larger $p_{c}$ will only increase $T_{F}$, and $T_{BG}$, but still maintain the relation $T_{F}$ $>$ $T_{BG}$. For acoustic phonon scattering at $T$ $>$ $T_{BG}$, $G$ is expected to follow a $\sim$ $T^{-1}$ law, resulting in an apparent metallic behavior\cite{MoS2Phonon}. The metallic phase in our devices does not stem solely from a quantum electronic mechanism.

The insulating phase, on the other hand, for $p$ $<$ $p_{c}$, is the expected behavior for a 2D system. While a strong localization effect at low carrier densities results in an insulating behavior, an alternate semiclassical percolation model can also explain this phenomenon\cite{SarmaPerc, MoS2WangMIT}. Density inhomogeneities induced by disorder are believed to block conductive paths in the channel at low carrier densities, leading to an insulating state due to percolation of carriers between the potential fluctuations. The similar values of $G_{c}$ = $O(e^2/h)$ expected for both the localization, and percolation mechanisms make it difficult to choose one to explain the insulating phase in our devices, as is the case for other 2D systems\cite{SarmaMIT}. The variation of $R_{c}$ \textit{vs.} $V_{TG}$, as a function of temperature is shown in Figure~\ref{fig4}(c). The variation of $R_{c}$ roughly resembles $G$, with $R_{c}$ increasing with reducing temperature for $V_{TG}$ $>$ -4 V, and varying weakly for large negative $V_{TG}$. The weak temperature dependence of $R_{c}$ when the channel is populated with holes ($V_{TG}$ $<$ -4 V) is consistent with prior reports of contact behavior in MoS$_2$ FETs\cite{MoS2Intrinsic,MoS2Graph}.

\begin{figure}
  \includegraphics{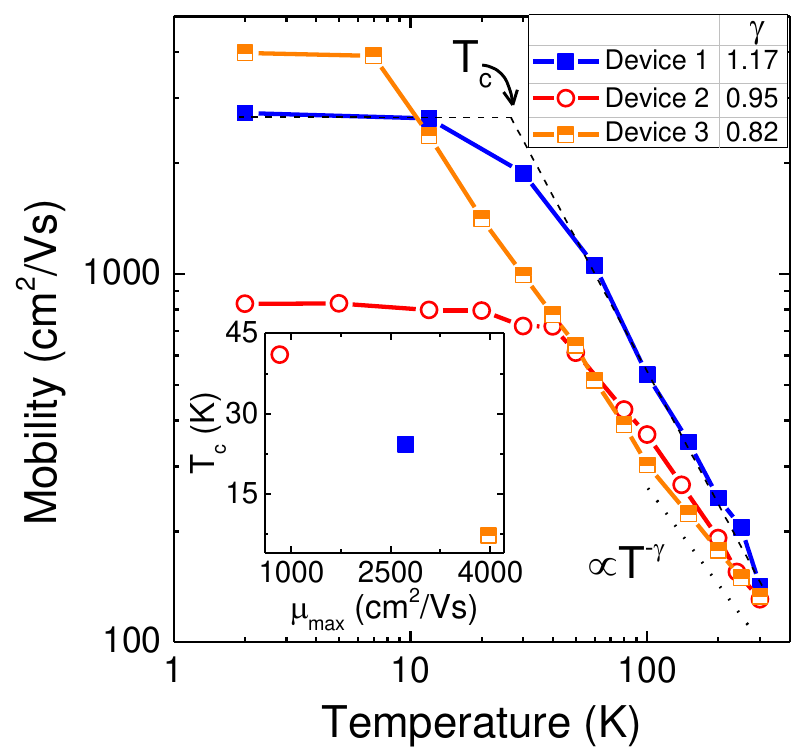}
  \caption{Temperature dependence of $\mu_{4pt}$ for three different WSe$_2$ devices. For $T$ $>$ 100 K, $\mu_{4pt}$ follows a power law trend, $\propto$ $T^{-\gamma}$, with the $\gamma$ values for the three devices shown in the inset table. At low-temperatures, $\mu_{4pt}$ saturates to $\mu_{max}$ $\sim$ 4,000 cm$^2$/Vs, limited by Coulomb scattering and/or defects in the WSe$_2$. The inset shows variation of $T_{c}$ with $\mu_{max}$.}
  \label{fig5}
\end{figure}

Finally, to determine the scattering mechanisms limiting hole transport in WSe$_2$, the variation of $\mu_{4pt}$ with temperature in the metallic regime is shown for three different three/four-layer WSe$_2$ FETs in Figure~\ref{fig5}. All three devices show a modest $\mu_{4pt}$ of around 140 cm$^2$/Vs at room temperature, which then increases rapidly with decreasing temperature. The variation in the high temperature regime (T $>$ 100 K) follows a power law dependence, $\mu_{4pt}$ $\propto$ $T^{-\gamma}$, with varying values of $\gamma$ (from $\sim$ 0.8 to 1.2) for the three samples measured. Acoustic phonon scattering is expected to result in $\gamma$ = 1, whereas $\gamma$ $>$ 1 is a signature of optical phonon scattering being the dominant scattering mechanism\cite{MoS2Phonon}. The values of $\gamma$ closer to 1 in our devices suggest that acoustic phonon scattering is the mobility limiting factor\cite{MoS2Acoustic}. Lowering of $\gamma$ below 1 has been attributed to homopolar phonon quenching by the top-gate dielectric, and/or the collective effect of multiple scattering mechanisms\cite{MoS2MIT, MoS2CVDMIT}. At low-temperatures, $\mu_{4pt}$ saturates to an upper limit ($\mu_{max}$), likely limited by Coulomb scattering, or defects\cite{MoS2Phonon,Mob2D}. A critical temperature ($T_{c}$) can be defined at the crossover of the two regimes of temperature dependence of $\mu_{4pt}$. There is a considerable variability in $\mu_{max}$ between samples ($\mu_{max}$ $\sim$ 800 cm$^2$/Vs to 4,000 cm$^2$/Vs), which varies inversely with $T_{c}$ ($\sim$ 40 K to 7 K), as shown in the inset of Figure~\ref{fig5}. The value of $T_{c}$ can be an indicator of sample quality, with cleaner samples transitioning to a Coulomb scattering dominated transport regime at lower temperatures\cite{Mob2D}. We also note that $\mu_{max}$ does not seem to depend on $\gamma$. We attribute the high values of $\mu_{max}$ in our devices to the cleaner top hBN-WSe$_2$ interface, where the holes reside at negative $V_{TG}$ values. The hole mobilities in our devices compare very well with recent reports of electron mobilities in hBN encapsulated WSe$_2$\cite{WSe2hBNEncap}, underlining the high material quality of WSe$_2$.

\section{Conclusion}

To summarize, we successfully used high work-function Pt electrodes to contact the valence band of WSe$_2$. Our structure with back-contacts, and an hBN top-gate dielectric provides a device design for optimized top-gated operation, resulting in stable Ohmic \textit{p}-type contacts without the need for any additional doping of the channel access regions. We observed saturating output characteristics, with signature of a back-gate tunable negative differential resistance. The Ohmic Pt contacts down to cryogenic temperatures enabled us to perform temperature dependent transport measurements which revealed a metal-insulator transition. The temperature dependence of mobility indicated a phonon dominated scattering mechanism at high-temperatures, with a crossover to Coulomb scattering at low-temperatures. Our findings highlight the significance of Pt as a \textit{p}-type contact for WSe$_2$ in order to study its intrinsic electrical properties. Moreover, the combination of our back-contact geometry, and an hBN top-gate dielectric provide a viable platform to explore the transport properties of other 2D materials, and their heterostructures.

\section{Materials and Methods}

\subsection{Device Fabrication}

The FETs are fabricated using commercially available sources of WSe$_2$ crystals (\hyperlink{www.hqgraphene.com}{HQ Graphene}, and \hyperlink{www.nanoscience.com}{nanoScience Instruments}). Both source materials result in devices with very similar characteristics. Individual flakes of WSe$_2$, and hBN are exfoliated onto 300 nm SiO$_2$/Si substrates, and three/four-layer WSe$_2$, and 15-20 nm hBN flakes are identified using optical contrast, Raman, and photoluminescence measurements. On a separate substrate, thin Cr/Pt (2 nm/8 nm) electrodes are patterned using a combination of e-beam lithography (EBL), e-beam metal evaporation (EBME), and lift-off. Using a silicone stamp spin-coated with a heat-release polymer\cite{OneDcontact}, we first ``pick-up" the hBN flake. A custom-built micromanipulator-microscope setup is then used to align, and ``pick-up" the WSe$_2$ using the hBN, resulting in an hBN/WSe$_2$ stack supported on the polymer. This stack is then aligned, and stamped on to the pre-patterned Cr/Pt electrodes, after which the polymer is washed away in acetone. This leaves the hBN/WSe$_2$ stack on the Cr/Pt electrodes. A 3 h 200$^{\circ}$C forming gas (1 Torr) anneal is performed to clean any remaining polymer residues. A local Pd top-gate (30 nm) is then patterned using EBL, EBME, and lift-off. Finally, thick Cr/Au (10 nm/80 nm) contact pads are patterned for electrical probing.

\subsection{Electrical Characterization}
Room temperature electrical measurements are done in ambient conditions, on a Cascade Summit probe station using an Agilent B1500A DC parameter analyzer. Temperature dependent measurements are done in a PPMS EverCool II Helium refrigerator. An Agilent B1500A is used for DC measurements, and an SR830 lock-in amplifier is used for the low-frequency lock-in measurements.

\begin{acknowledgement}
This work was supported in part by NRI SWAN, Intel Corp., and the NSF NNIN program.
\end{acknowledgement}

\subsection{Supporting Information Available}
S1: Raman and photoluminescence characteristics, S2: Transfer characteristics and top-gate leakage, S3: 2-point and 4-point measurement scheme, S4: Top-contacts \textit{vs} back-contacts, S5: Hysteresis in output characteristics,  S6: Metal-insulator transition, S7: Temperature scales for MIT.


\setcounter{figure}{0}
\makeatletter 
\renewcommand{\thefigure}{S\arabic{figure}}

\begin{center}
{\LARGE \bf Supporting Information}
\end{center}

\section{S1: Raman and photoluminescence characteristics}\label{Sec1}

\begin{figure*}
  \includegraphics{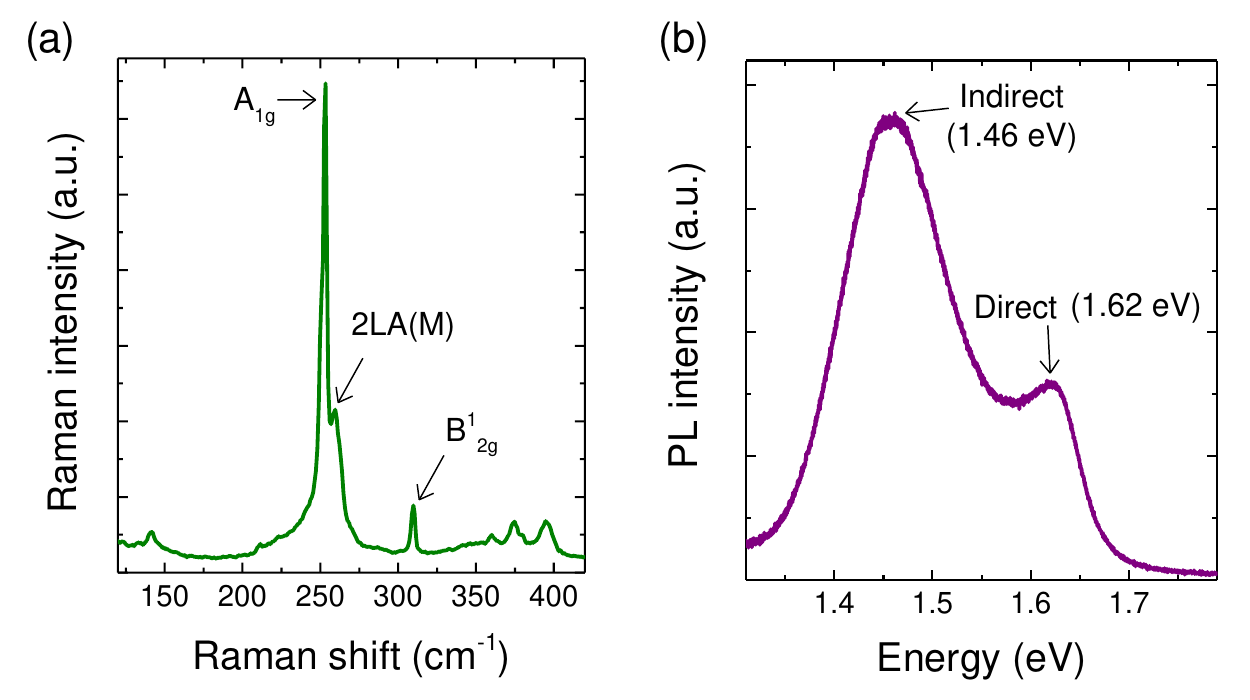}
  \caption{(a) Raman spectrum of a typical exfoliated three/four-layer WSe$_2$ flake, with the peaks labeled. (b) The PL spectrum shows two peaks arising from the direct, and indirect transitions.}
  \label{S0}
\end{figure*}

Figure~\ref{S0} shows the Raman and photoluminescence (PL) spectra of a typical three/four-layer exfoliated WSe$_2$ flake, measured using a 532 nm laser excitation. The sharp Raman peaks shown in Figure~\ref{S0}(a) reflect the good material quality of the flakes used in this work. The full width at half maximum for the A$_\text{1g}$ peak is $<$ 3 cm$^{-1}$ indicating high in-plane crystallinity\cite{WSe2Raman}. This is further corroborated by the PL spectrum shown in Figure~\ref{S0}(b). The two distinct peaks correspond to the direct bap transition at 1.62 eV, and the indirect gap transition at 1.46 eV\cite{WSe2Kyoung}. 

\section{S2: Transfer characteristics and top-gate leakage}\label{Sec2}

\begin{figure*}
  \includegraphics{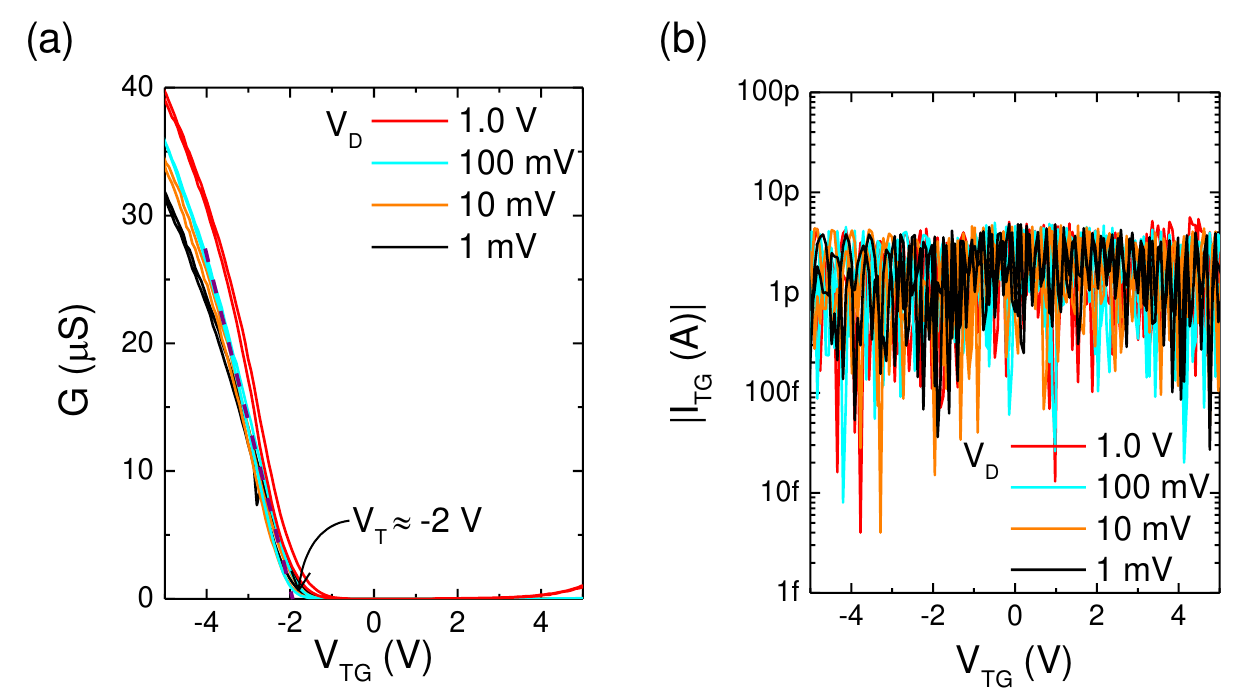}
  \caption{(a) Top-gated transfer characteristics of the FET on a linear scale. (b) The top-gate leakage current stays at the noise-floor throughout the measurement range.}
  \label{S1}
\end{figure*}

Figure~\ref{S1}(a) shows the top-gated transfer characteristics of the FET plotted on a linear scale. A threshold voltage ($V_{T}$) $\sim$ -2 V can be extracted by extrapolating the linear region to the top-gate voltage ($V_{TG}$) axis. The top-gate leakage is shown in Figure~\ref{S1}(b), which stays at the noise-floor of the measurement setup throughout the range of $V_{TG}$ probed.

\section{S3: 2-point and 4-point measurement scheme}\label{Sec3}

\begin{figure*}
  \includegraphics{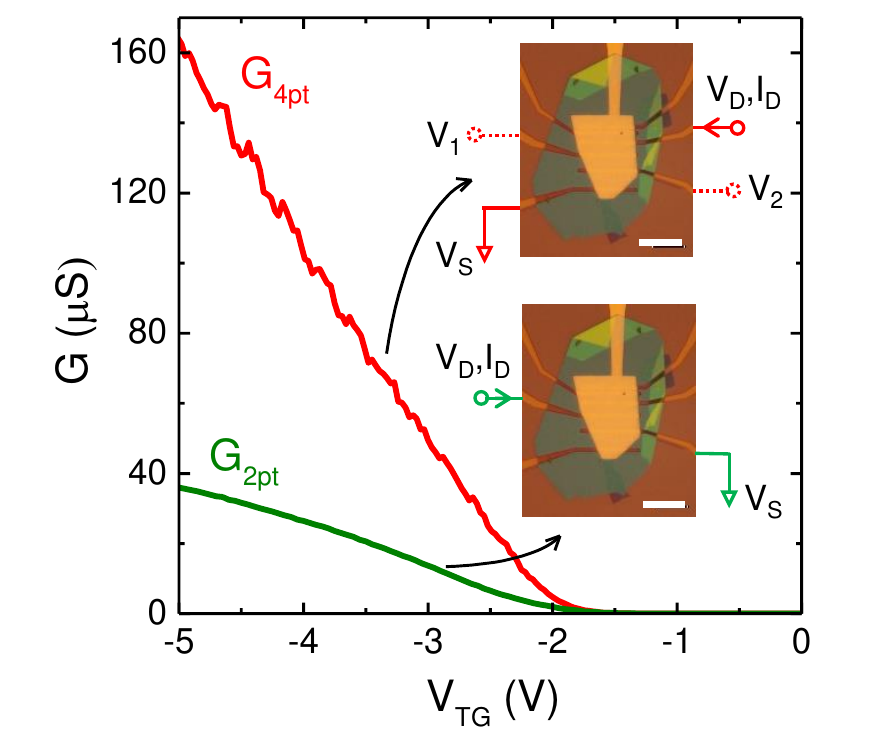}
  \caption{The biasing scheme used for measurement of $G_{2pt}$ and $G_{4pt}$. The scale bars are 10 $\mu$m.}
  \label{S2}
\end{figure*}

Figure~\ref{S2} shows the biasing scheme used for measurement of the 2-point conductance ($G_{2pt}$), and intrinsic, 4-point conductance ($G_{4pt}$). For measuring $G_{2pt}$, a pair of adjacent contacts are used as the source and drain, and $G_{2pt}$ = $I_{D}$/$V_{D}$. To measure $G_{4pt}$ between the same two contacts, an outer pair of contacts are chosen as the source and drain. The voltage drop between the original pair of contacts ($V_1$-$V_2$) is then measured with a current ($I_{D}$) flowing through the outer pair of contacts. Now, $G_{4pt}$ = $I_{D}$/($V_1$-$V_2$).

\section{S4: Top-contacts \textit{vs} back-contacts}\label{Sec4}

\begin{figure*}
  \includegraphics{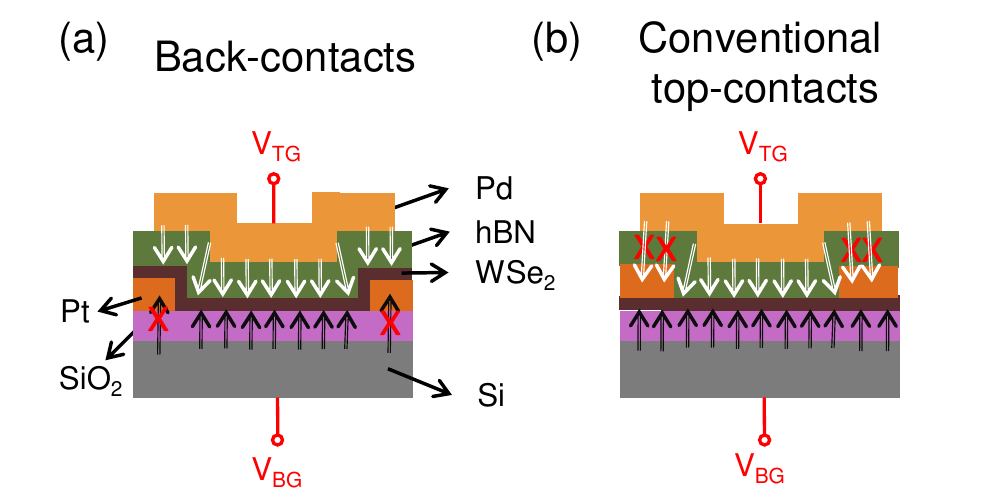}
  \caption{(a) Schematic of our dual-gated FET structure with back-contacts. (b) Schematic of conventional TMD FETs with top-contacts.}
  \label{S3}
\end{figure*}

Figures~\ref{S3}(a), and (b) show the schematics of our FET structure with back-contacts, and a conventional FET with top-contacts, respectively. The field lines from the top-gate are represented by white arrows, and the field lines from the back-gate by black arrows. In Figure~\ref{S3}(a), the top-gate is able to modulate the channel, the contact regions, and the channel access regions. Field lines from the back-gate, however, are screened out by the Pt back-contacts (shown by red crosses). Back-gated transfer characteristics are therefore severely contact resistance dominated. In contrast, for a conventional top-contact geometry (Figure~\ref{S3}(b)), the top-gate is unable to modulate the contact regions due to screening by the top-contact electrodes. Consequently, the top-gated transfer characteristics display a similar series resistance limited behavior\cite{MoS2Kis,MoS2Int,WSe2Fang}. Placing contacts underneath the flake optimizes our structure for top-gated operation.

\section{S5: Hysteresis in output characteristics}\label{Sec5}

\begin{figure*}
  \includegraphics{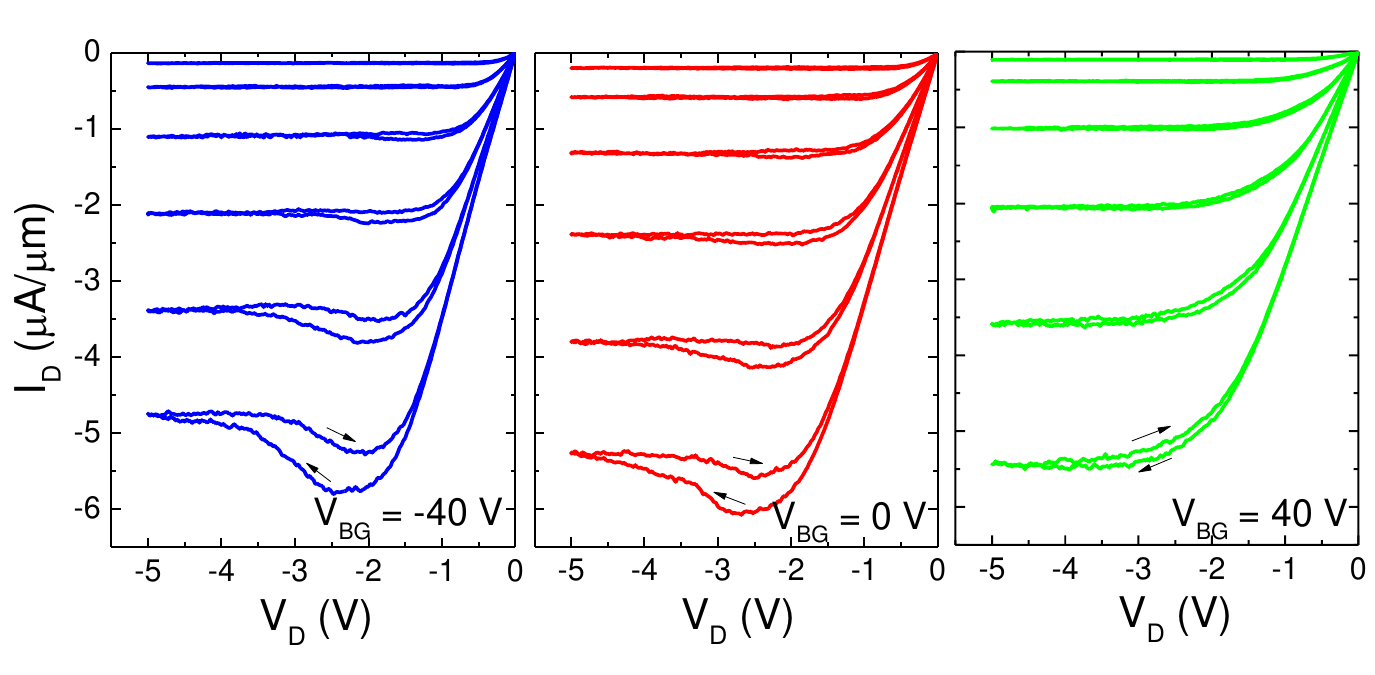}
  \caption{Forward and reverse $I_{D}$-$V_{D}$ sweeps show hysteresis at the NDR point, which reduces with the application of a positive $V_{BG}$.}
  \label{S3b}
\end{figure*}

Figure~\ref{S3b} shows the forward and reverse sweeps of the FET output characteristics at different values of $V_{BG}$. A negative differential resistance (NDR), and a corresponding hysteresis near the NDR region are apparent. The hysteresis is negligible both for low $V_{D}$, and large negative $V_{D}$. Both the NDR, and hysteresis increase (decrease) for negative (positive) $V_{BG}$, suggestive of hot carrier trapping at the WSe$_2$-SiO$_2$ interface\cite{WSe2Kyoung}. It is to be noted that negligible carrier trapping at low $V_{D}$ indicates clean interfaces for low energy holes. Only hot holes generated at a sufficiently large negative $V_{D}$ can overcome the potential barrier to enter a trap state.

\section{S6: Metal-insulator transition}\label{Sec6}

\begin{figure*}
  \includegraphics{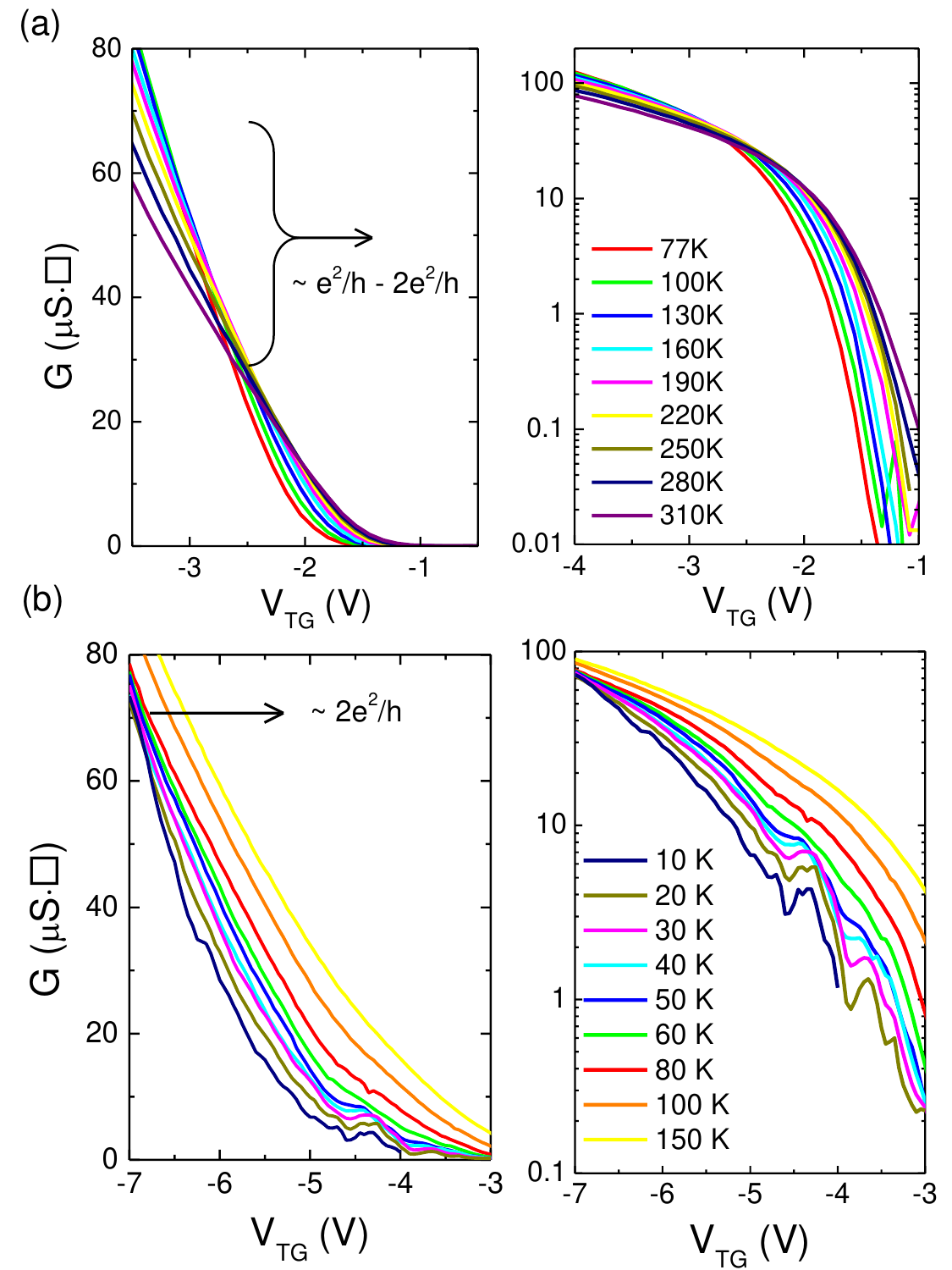}
  \caption{Conductivity temperature dependence for two other devices. A clear insulating state for low $V_{TG}$ is apparent for both devices in the log scale (right-top, and right-bottom panels)}
  \label{S4}
\end{figure*}

Figure~\ref{S4} shows temperature dependence of $G$ for two other representative devices. The data for the device in Figure~\ref{S4}(a) shows a broad MIT crossover in the range of $G_{c}$ $\sim$ $e^2/h$\textemdash$2e^2/h$. A clear insulating state is observed for $V_{TG}$ $>$ -2.5 V. The device in Figure~\ref{S4}(b) shows a crossover point developing at $V_{TG}$ = -7 V, at $G_{c}$ $\sim$ 2$e^2/h$, similar to a prior report on monolayer WSe$_2$\cite{WSe2mono}. The oscillatory behavior of $G$ at low-temperatures is indicative of disorder, and charge puddles which cause mesoscopic fluctuations with varying $V_{TG}$\cite{MoS2Ghosh}.

\section{S7: Temperature scales for MIT}\label{Sec7}

\begin{figure*}
  \includegraphics{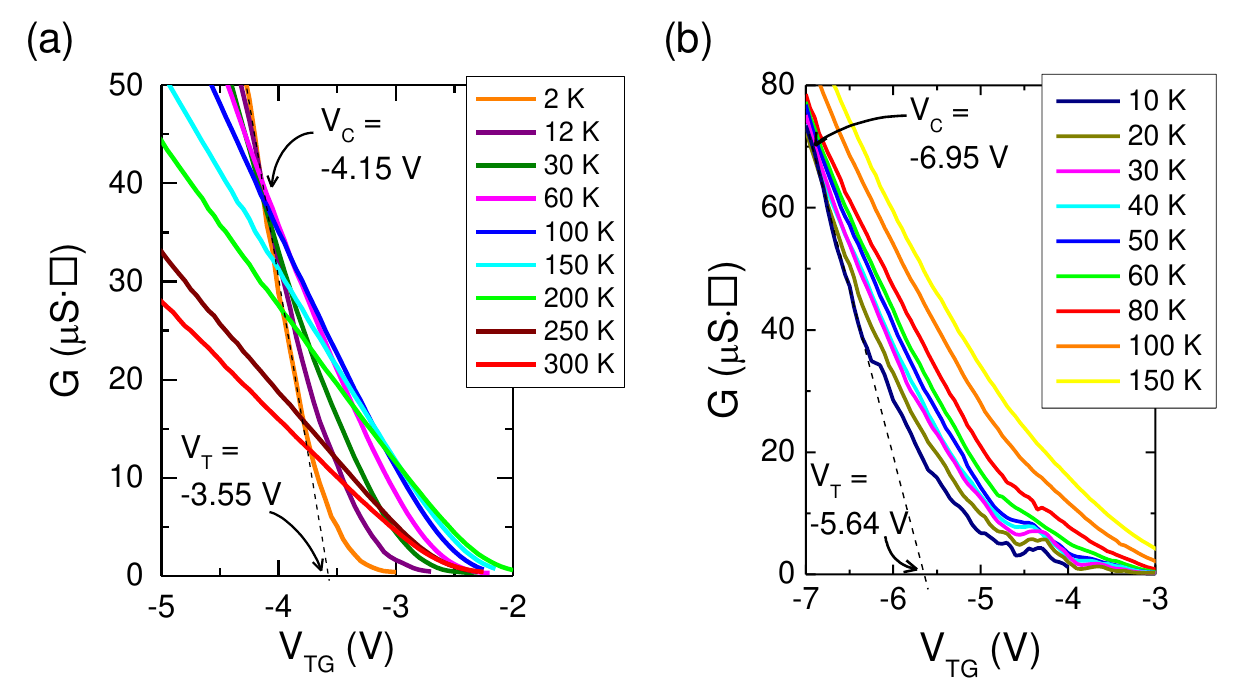}
  \caption{Close up of the MIT crossover for two devices. The device in (a) has $\mu_{4pt}$ = 2,600 cm$^2$/Vs, and the device in (b) has $\mu_{4pt}$ = 800 cm$^2$/Vs.}
  \label{S5}
\end{figure*}

Figure~\ref{S5} shows the MIT crossover for two WSe$_2$ devices, with the crossover voltage ($V_{C}$), and $V_{T}$ marked. The carrier density at crossover ($p_{c}$) can then be calculated using ${{p_{{c}} = C_{{TG}}|(V_{{C}}-V_{{T}})|}}/e$, where $e$ is the electron charge. Three different temperature scales can then be defined according to ref.\cite{SarmaMIT} as - the electron temperature scale defined by the Fermi temperature ($T_{F}$), the phonon temperature scale defined by the Bloch-Gr{\"u}neisen temperature ($T_{BG}$), and the disorder temperature scale defined by the Dingle temperature ($T_{D}$), as follows:

\begin{equation}
k_{B}{T}_{F} = E_{F} = \frac{{\hbar^2}{k_{F}^2}}{2m} = \frac{{\hbar^2}{\pi{p_{c}}}}{m}
\label{eqs1}
\tag{S1}
\end{equation}

\begin{equation}
k_{B}{T}_{BG} = 2\hbar{k_{F}}v_{ph} = 2\hbar{v_{ph}}\sqrt{2{\pi}p_{c}}
\label{eqs2}
\tag{S2}
\end{equation}

\begin{equation}
k_{B}{T}_{D} = \Gamma = \frac{\hbar}{2}\Big(\frac{e}{m\mu}\Big)
\label{eqs3}
\tag{S3}
\end{equation}

Here, $E_{F}$ is the Fermi energy, $k_{F} = \sqrt{2\pi{p_{c}}}$ is the Fermi wave vector, $m$ is the carrier effective mass, $v_{ph}$ is the phonon velocity, $\Gamma$ is the impurity-scattering induced level broadening, and $\mu$ is the carrier mobility. We assume a spin-degeneracy of 2, and a valley degeneracy of 1 for relating $k_{F}$ and $p_{c}$, $m = 0.5m_{e}$, where $m_{e}$ is the free electron mass, and $v_{ph}$ = 6$\times{10^3}$ m/s is the phonon velocity\cite{MoS2Phonon,TMDBands}. $k_B$ is the Boltzmann constant, and $\hbar$ is the reduced Planck constant.

For the device in Figure~\ref{S5}(a), $p_\text{c}$ = 5.3$\times{10^{11}}$/cm$^2$, and $\mu_{4pt}$ = 2,600 cm$^2$/Vs at 2 K. We therefore calculate $T_{F}$ = 29 K, $T_{BG}$ = 16 K, and $T_{D}$ = 5.0 K at the crossover point.  

For the device in Figure~\ref{S5}(b), with $p_{c}$ = 1.1$\times{10^{12}}$/cm$^2$, and $\mu_{4pt}$ = 800 cm$^2$/Vs at 10 K, we calculate $T_{F}$ = 62 K, $T_{BG}$ = 24 K, and $T_{D}$ = 17 K. We find $T_{F}$ $>$ $T_{BG}$ even for this device, indicating a phonon scattering effect to be the cause of the metallic behavior.

It is to be noted that given the lack of Hall measurements, and the uncertainty in $V_{T}$, the value of $p_{c}$ used for calculating $T_{F}$, and $T_{BG}$ could be underestimated. However, even when considering the extreme case of $V_{T}$ = 0 V for the device in Figure~\ref{S5}(a), we obtain $p_\text{c}$ = 3.7$\times{10^{12}}$/cm$^2$, which increases $T_{F}$, and $T_{BG}$ to $T_{F}$ = 200 K,  and $T_{BG}$ = 42 K. Since $T_{D}$ depends only on the mobility, it remains unchanged. The relation $T_{F}$ $>$ $T_{BG}$ is maintained even in this case.

\bibliography{hemacp_WSe2FETs}

\end{document}